\documentclass[prd,showpacs,10pt]{revtex4}
\usepackage{graphicx}
\usepackage{bm}

\newcommand{\square}{\kern1pt\vbox{\hrule height 1.2pt\hbox{\vrule
width 1.2pt\hskip 3pt
\vbox{\vskip 6pt}\hskip 3pt\vrule width 0.6pt}\hrule
height 0.6pt}\kern1pt}
\newcommand{\beq}{\begin{equation}}
\newcommand{\beqn}{\begin{eqnarray}}
\newcommand{\eeq}{\end{equation}}
\newcommand{\eeqn}{\end{eqnarray}}

\begin{document}

\title{Cosmic No Hair for Braneworlds with a Bulk Dilaton Field}
\author{James E. Lidsey and David Seery}
\affiliation{Astronomy Unit, School of Mathematical Sciences, 
Queen Mary, University of London, Mile End Road,
London, E1 4NS, UK}


\begin{abstract}
Braneworld cosmology supported by a bulk scalar 
field with an exponential potential is developed. A general class 
of separable backgrounds for both single and two--brane systems
is derived, where the bulk metric components 
are given by products of world-volume and bulk 
coordinates and the world--volumes represent any anisotropic and 
inhomogeneous solution to an effective four--dimensional Brans--Dicke 
theory of gravity. We deduce a cosmic no hair theorem for 
all ever expanding, spatially homogeneous Bianchi world-volumes and 
find that the spatially flat and isotropic inflationary scaling solution 
represents a late-time attractor when the bulk potential is 
sufficiently flat. The dependence of this result on the 
separable nature of the bulk metric is investigated 
by applying the techniques of Hamilton-Jacobi theory to 
five-dimensional Einstein gravity. We employ the 
spatial gradient expansion method to determine the 
asymptotic form of the bulk metric up to third-order in spatial gradients. 
It is found that the condition for the separable form of the metric 
to represent the attractor of the system is precisely the same as that 
for the four-dimensional world-volume to isotropize. We also derive  
the fourth--order contribution to the Hamilton-Jacobi 
generating functional. Finally, we conclude by placing 
our results within the context of the holographic 
approach to braneworld cosmology. 
 \end{abstract}

\vskip 1pc 
\maketitle

\section{Introduction}

One of the most striking features of our observable universe is that on
sufficiently large scales it is very nearly 
spatially isotropic and homogeneous. The inflationary paradigm provides an 
attractive, dynamical mechanism for the universe to evolve 
into such a symmetric state from a potentially wide class of 
anisotropic and inhomogeneous initial conditions \cite{wald,km,othernohair}. 
Such a feature is generally referred to as `cosmic no hair'. 
Braneworld cosmology, motivated by string/M--theoretic considerations
\cite{stringreview}, 
has received considerable attention in recent years. (For reviews, 
see, e.g., Refs. \cite{branereview}). In this 
scenario our four--dimensional universe is viewed as a co--dimension one brane 
embedded in a higher--dimensional `bulk' space. It is important, therefore, 
to investigate the isotropization of the universe within 
the context of braneworld inflation \cite{nohairbrane,5danisotropic}. 

To date, however, progress in this direction has been hindered by 
our lack of knowledge of the geometry of the bulk space. 
In the Randall--Sundrum (RS) scenario \cite{RSbrane}, for example, 
our braneworld is embedded in a five--dimensional Einstein space 
sourced by a negative cosmological constant. In particular,  
a spatially isotropic braneworld propagates in 
five--dimensional Anti-de Sitter (AdS)--Schwarzschild space. 
On the other hand, due to the complexity of the 
field equations, very few exact anisotropic (or inhomogeneous) 
solutions to the five--dimensional bulk Einstein equations  
have been found \cite{5danisotropic}. 

A natural and well-motivated extension of the RS scenario is to include 
one or more scalar fields in the bulk action.
In this paper we consider a five--dimensional action for gravity 
coupled to two scalar fields $\{ \varphi , \sigma \}$: 
\begin{equation}
\label{action5}
S_5= \int_{{\cal{M}}_5} d^5 x \sqrt{-\hat{g}} \left[ \hat{R} - \frac{1}{2} 
( \hat{\nabla} \varphi )^2 - \frac{1}{2} 
e^{-b\varphi} ( \hat{\nabla} \sigma )^2 - V(\varphi )
 \right] +
 \sum_{i=1}^2 \int_{{\cal{M}}^{(i)}_4} 
d^4x \sqrt{-g_i} T_i(\varphi ) , 
\end{equation}
where the `dilaton' field $\varphi$ self--interacts through a potential 
$V(\varphi )$ and is coupled to the branes through its 
brane potentials $T_i (\varphi )$ that are localized to the 
branes \footnote{The bulk spacetime metric has signature $(-,+,+,+,+)$. 
Upper case, Latin indices vary from $(0, \ldots , 4)$ and 
lower case, Greek indices span
$(\mu = 0, 1, 2, 3)$. A circumflex accent denotes geometric quantities 
compatible with the five--dimensional metric $\hat{g}_{AB}$.
The Ricci curvature scalar of the bulk spacetime ${\cal{M}}_5$ is denoted by 
$\hat{R}$, $\hat{g} \equiv {\rm det} \hat{g}_{AB}$ 
and $g^{(i)} \equiv {\rm det} g^{(i)}_{\mu\nu}$.}.  
In the case where the bulk dimension has the topology $S^1/{\rm Z}_2$, 
we may consider two `end-of-the-world' branes 
located at the orbifold fixed points. 
The metrics induced on these four--dimensional hypersurfaces 
are defined by $g^{(1)}_{\mu\nu} \equiv 
\hat{g}_{\mu\nu} (y=0)$ and $g^{(2)}_{\mu\nu} \equiv 
\hat{g}_{\mu\nu} (y=\pi )$, respectively, where the fifth dimension is 
parametrized by the coordinate $y$. If 
the bulk dimension is not periodic, we may consider the case of 
a single brane by specifying $T_1 \ne 0$ and $T_2 = 0$ 
such that the ${\rm Z}_2$ symmetry is still respected across the brane. 
The constant $b$ determines the coupling between the dilaton and 
massless `axion' field, $\sigma$. 

We will focus on the class of exponential self--interaction 
potentials: 
\begin{equation}
\label{potential5}
V (\varphi ) = V_0 e^{-q\varphi} , \quad T_i (\varphi ) = 
\mu_i e^{-q\varphi/2}  ,
\end{equation}
where $\{ q, V_0 , \mu_i \}$ are constants. 
An action of the form (\ref{action5})--(\ref{potential5})
is well motivated from  a number of perspectives. 
When $q=2$, $b=1$ and $\mu_1 = 
-\mu_2 = \sqrt{6V_0}$, it represents a consistent 
truncation of Ho\v{r}ava--Witten theory \cite{HW}
compactified on a Calabi--Yau three--fold, where the 
dilaton represents the breathing mode of the 
Calabi--Yau space and the axion arises from the 
universal hypermultiplet \cite{lukas}. The potential 
is generated by the non--trivial flux of the four--form field 
strength on four--cycles of the Calabi--Yau space. 
In the absence of an axion field, 
action (\ref{action5})--(\ref{potential5}) also follows from 
the toroidal, Kaluza--Klein compactification of the $(5+m)$--dimensional 
RS model, where the dilaton again represents 
the breathing mode of the internal dimensions and 
the exponential coupling is given by $q= \sqrt{2m/[3(m+3)]}$ 
\cite{kt}.  

Recently, cosmological solutions with a 
single brane \cite{koyama} and a two--brane configuration \cite{mc} 
were found for action (\ref{action5})--(\ref{potential5}) with a
vanishing axion field when the world--volume of the branes is represented by 
the spatially flat and isotropic Friedmann--Robertson--Walker 
(FRW) metric. In both cases, the five--dimensional bulk solution has the form 
\begin{equation}
\label{scalingsol}
ds^2 = n^2(y) \left[ -dt^2 +t^{4/(3q^2)} \delta_{ij} dx^idx^j 
+t^2dy^2 \right]   ,
\end{equation}
where $n \propto e^{2\psi/3q^2}$, $e^{-q\varphi} \propto t^{-2}e^{-2\psi}$
and $\psi=\psi(y)$ is a function of $y$ that is determined by solving 
the field equations. This 
solution may be interpreted by an observer confined to the 
brane as a power--law  inflationary cosmology 
for $q^2< 2/3$ and inflation proceeds as the inter-brane distance 
increases. Solution (\ref{scalingsol}) is `separable', in the 
sense that physical quantities are represented as products 
of functions of $t$ and $y$. This corresponds physically to the case
where there is no net propagation of scalar waves in the bulk. 

The solution ({\ref{scalingsol}) also represents a scaling 
solution, since the Hubble parameter on the brane and the dilaton's 
kinetic energy scale at the same rate. 
In general, scaling solutions play an important role in 
cosmology. They establish the asymptotic 
behaviour of a particular cosmology as well as determining its stability 
properties. Moreover, the attractive nature of scaling solutions 
provides a dynamical framework where the initial conditions for any 
subsequent cosmological evolution can be well--defined. 
Mukohyama and Coley \cite{mc} have shown that in the two-brane scenario 
the scaling solution (\ref{scalingsol}) is stable against homogeneous 
linear metric perturbations and have also found that 
it represents the late--time attractor for a general 
FRW world--volume. Recently, however, an unstable mode 
has been identified in the case where the position of 
one of the branes is perturbed without producing a corresponding 
metric perturbation and this can result in a brane collision \cite {leeper}.

The purpose of the present work is to investigate the implications 
of relaxing the separable ansatz discussed above 
as well as the assumption that the world-volume corresponds to an isotropic 
FRW metric. In Section II, we first 
consider the case of an arbitrary world--volume metric and 
show that, in general, the effective dynamics on the brane 
can be described by a Brans--Dicke scalar--tensor theory of 
gravity, where the coupling between the four--dimensional 
dilaton and graviton degrees of freedom is 
given by $\omega = 2/q^2$. This leads us to deduce a 
cosmic no hair theorem for this class of models. In particular, 
we find that all initially expanding, spatially homogeneous 
Bianchi type I--VIII models will isotropize into 
the future when $q^2< 2/3$. This implies that there is an open set of 
Bianchi models for which the scaling solution (\ref{scalingsol}) is an 
attractor at late times.  

In Section III, we investigate the consequences of relaxing the 
separable ansatz. This yields insight into the 
nature of scalar wave propagation in the 
bulk. A full analysis would involve a search for inhomogeneous solutions 
to the field equations of action (\ref{action5})--(\ref{potential5}). 
A powerful way of integrating 
Einstein's equations is to apply the techniques of 
Hamilton--Jacobi theory to general relativity. 
A systematic and non--linear scheme for solving the 
Hamilton--Jacobi equation has been developed by Salopek and 
collaborators by employing a spatial gradient expansion \cite{sal1,sal2,sal3}.  
We employ this method in a five--dimensional context to solve 
the evolution equation for the four--metric to third--order in spatial 
gradients. When the dilaton field is homogeneous on the 
world--volume, $\varphi =\varphi (y)$, it 
is found that the separable solution represents 
the attractor of the system when $q^2< 2/3$. 
Combining the results of 
Sections II and III therefore provides strong evidence that the scaling 
solution (\ref{scalingsol}) represents an attractor for $q^2 < 2/3$. 
Finally, we conclude in Section IV by placing our results within 
the context of the holographic approach to braneworld cosmology motivated by 
the AdS/CFT correspondence. 

\section{Separable Braneworlds}

\subsection{Field Equations}

The bulk field equations derived by extremizing action 
(\ref{action5})--(\ref{potential5}) are given by 
\begin{eqnarray}
\nonumber 
\hat{G}_{AB} = \frac{1}{2} \hat{\nabla}_A \varphi \hat{\nabla}_B 
\varphi + \frac{1}{2} e^{-b\varphi} \hat{\nabla}_A \sigma 
\hat{\nabla}_B \sigma - \hat{g}_{AB} \left[ 
\frac{1}{4} \left( \hat{\nabla} \varphi \right)^2 + \frac{1}{4}
e^{-b\varphi} \left( \hat{\nabla}\sigma \right)^2 
+ \frac{V_0}{2} e^{-q\varphi} \right] 
\\
\label{einsteinfield5}
+\hat{g}_{\mu A} \hat{g}_{\nu B} e^{-q\varphi /2} 
\left[ \mu_1 \delta (y) g_{(1)}^{\mu \nu} \sqrt{\frac{g_{(1)}}{\hat{g}}}
+ \mu_2 \delta(y-\pi ) g^{\mu\nu}_{(2)} \sqrt{\frac{g_{(2)}}{\hat{g}}}
\right]
\\
\label{scalarfield5}
\hat{\square} \varphi = - \frac{b}{2} e^{-b\varphi} 
\left( \hat{\nabla} \sigma \right)^2 - q V_0 
e^{-q\varphi} + q e^{-q\varphi /2} \left[ \mu_1
\delta (y) \sqrt{\frac{g_{(1)}}{\hat{g}}} + \mu_2 
\delta (y-\pi ) \sqrt{\frac{g_{(2)}}{\hat{g}}} \right]
\\
\label{sigmafield5}
\hat{\nabla}_A \left( e^{-b\varphi}
\sqrt{-\hat{g}} \hat{g}^{AB} \hat{\nabla}_B \sigma \right) = 0   .
\end{eqnarray}

In this Section, we assume a bulk metric of the general form
\begin{equation}
\label{5dmetric}
d\hat{s}^2 = H^m \left( f_{\mu\nu} dx^{\mu}dx^{\nu} + e^{2 \beta} dy^2
\right) 
\end{equation}
where the world--volume metric is represented by 
$f_{\mu\nu} = f_{\mu\nu} (x^{\rho})$, $H=H (y)$ denotes the warp factor, 
$\beta =\beta (x)$ may be interpreted in four dimensions as
a `radion' field and $m \equiv 4/(3q^2 -2)$. 
The components of the five--dimensional 
Ricci tensor compatible with the metric (\ref{5dmetric}) are
then given by 
\begin{eqnarray}
\label{ricci5munu}
\hat{R}_{\mu\nu} & = & R_{\mu\nu} - \nabla_{\mu\nu} \beta 
-\nabla_{\mu}\beta \nabla_{\nu} \beta + 
\frac{m}{2} e^{-2\beta} \left[ 
\left( 1-\frac{3m}{2} \right) \frac{H'^2}{H^2} -\frac{H''}{H} 
\right] f_{\mu\nu}
\\
\label{riccimuy}
\hat{R}_{\mu y}
& = & \frac{3m}{2} \frac{H'}{H} \nabla_{\mu} \beta
\\
\label{ricci5yy}
\hat{R}_{yy} & = & - e^{2\beta} \left[ \square \beta +\left( 
\nabla \beta \right)^2 \right] +2m \left( \frac{H'^2}{H^2}
- \frac{H''}{H} \right)    ,
\end{eqnarray}
where a prime denotes differentiation with respect to $y$.

\subsection{Separable Branes}

A separable solution between the world--volume and bulk 
coordinates can be found by assuming the {\em ansatz}
$\varphi = \varphi_1 (x) + \varphi_2 (y)$ and $\sigma = \sigma (y) $
and specifying $b= 2/q$. 
The $(\mu y)$--components of the Einstein field equations
(\ref{einsteinfield5}) are then solved directly by 
\begin{equation}
\label{offdiagsolve}
\varphi_1 = \frac{2}{q} \beta , \qquad \varphi_2 = 
\frac{6q}{3q^2-2} \ln H
\end{equation}
and the axion field equation (\ref{sigmafield5}) admits the first integral
\begin{equation}
\sigma' = \sigma_0 H^{6/(3q^2-2)}
\end{equation}
for an arbitrary constant $\sigma_0$.

The $(\mu\nu )$-- and $(yy)$-components of the field 
equations (\ref{einsteinfield5}) then reduce to 
\begin{eqnarray}
\label{reduce1}
G_{\mu\nu} - \nabla_{\mu\nu} \beta+f_{\mu\nu} \square \beta
- \left( 1+\frac{2}{q^2} \right) \nabla_{\mu} \beta \nabla_{\nu} \beta
+ \left( 1+\frac{1}{q^2} \right) f_{\mu\nu} \left( 
\nabla \beta \right)^2 
\nonumber \\
= - \frac{1}{4} \sigma_0^2 e^{-[2+2(b/q)]\beta} f_{\mu\nu}
+f_{\mu\nu}e^{-2\beta} 
\left[ -\frac{6}{3q^2-2} \frac{H''}{H} + \frac{3(3q^2-8)}{(3q^2-2)^2}
\frac{H'^2}{H^2} - \frac{V_0}{2H^2} 
+ \frac{1}{H} \left[ \mu_1 \delta(y) +\mu_2\delta (y-\pi) \right] \right]
\end{eqnarray}
and 
\begin{equation}
\label{reduce2}
\frac{R}{2} - \frac{1}{q^2} \left( \nabla \beta \right)^2 
=- \frac{\sigma_0^2}{4} e^{-[2+(2b/q)]\beta} 
+e^{-2\beta} \left[ \frac{3(8-3q^2)}{(3q^2-2)^2} \frac{H'^2}{H^2}
+ \frac{V_0}{2H^2} \right]  ,
\end{equation}
respectively, whereas the scalar field equation (\ref{scalarfield5}) 
takes the form  
\begin{eqnarray}
\label{reduce3}
e^{-\beta}\square   e^{\beta}  
 + \frac{bq}{4} \sigma^2_0 e^{-[2+(2b/q)]\beta}
= - \frac{q^2}{2} e^{-2\beta} \left[ 
\frac{6}{3q^2-2} \frac{H''}{H} + \frac{6(8-3q^2)}{(3q^2-2)^2}
\frac{H'^2}{H^2} + \frac{V_0}{H^2} - \frac{1}{H} 
\left( \mu_1 \delta (y) +\mu_2 \delta (y-\pi ) \right) \right]   .
\end{eqnarray}

A crucial property of Eqs. (\ref{reduce1})--(\ref{reduce3}) 
is that the terms contained within the square brackets 
are {\em independent} of the world--volume coordinates, $x^{\mu}$.
We therefore define three separation constants, $c_i$: 
\begin{eqnarray}
\label{c1define}
-\frac{6}{3q^2-2} \frac{H''}{H} 
+ \frac{3(3q^2-8)}{(3q^2-2)^2} \frac{H'^2}{H^2} 
- \frac{V_0}{2H^2} + \frac{1}{H} \left[ \mu_1 \delta (y) 
+\mu_2 \delta (y-\pi ) \right] \equiv c_1
\\
\label{c2define}
\frac{3(8-3q^2)}{(3q^2-2)^2} \frac{H'^2}{H^2} 
+ \frac{V_0}{2H^2} \equiv c_2
\\
\label{c3define}
-\frac{q^2}{2} \left[ \frac{6}{3q^2-2} \frac{H''}{H} 
+ \frac{6(8-3q^2)}{(3q^2-2)^2} \frac{H'^2}{H^2}
+ \frac{V_0}{H^2} - 
\frac{1}{H} \left[ \mu_1 \delta (y) + \mu_2 \delta (y-\pi) \right] \right]
\equiv c_3   .
\end{eqnarray}
This implies that 
Eqs. (\ref{reduce1})--(\ref{reduce3}) simplify to 
\begin{equation}
\label{efereduce1}
G_{\mu\nu} - \nabla_{\mu\nu} \beta+f_{\mu\nu} \square \beta
- \left( 1+\frac{2}{q^2} \right) \nabla_{\mu} \beta \nabla_{\nu} \beta
+ \left( 1+\frac{1}{q^2} \right) f_{\mu\nu} \left( 
\nabla \beta \right)^2 = 
- \frac{1}{4} \sigma_0^2 e^{-[2+2(b/q)]\beta} f_{\mu\nu}
+ c_1 f_{\mu\nu}e^{-2\beta} 
\end{equation}
\begin{equation}
\label{efereduce2}
\frac{R}{2} - \frac{1}{q^2} \left( \nabla \beta \right)^2 
=- \frac{\sigma_0^2}{4} e^{-[2+(2b/q)]\beta} 
+c_2 e^{-2\beta} 
\end{equation}
\begin{equation}
\label{efereduce3}
\square \beta + \left( \nabla \beta \right)^2 = - \frac{bq}{4} \sigma_0^2
e^{-[2+(2b/q)] \beta} + c_3 e^{-2\beta}   .
\end{equation}

However, the constants $c_i$ are not independent. 
Substituting the trace of Eq. (\ref{efereduce1}) into 
Eq. (\ref{efereduce2}) and comparing with Eq. (\ref{efereduce3}) 
implies that 
\begin{equation}
\label{cconstraint1}
3c_3 = 4c_1 +2c_2
\end{equation}
and subtracting 
Eq. (\ref{c1define}) from Eq. (\ref{c3define}) 
and comparing with Eq. (\ref{c2define}) implies that 
\begin{equation}
\label{cconstraint2}
c_1=c_2 +\frac{2}{q^2} c_3   .
\end{equation}
Eqs. (\ref{cconstraint1}) and (\ref{cconstraint2}) 
then imply that 
\begin{equation}
\label{cconstraint3}
6c_1 = \left( 3+\frac{4}{q^2} \right) c_3  .
\end{equation}

It may now be verified directly that 
Eqs. (\ref{efereduce1})--(\ref{efereduce3}) follow by  
extremizing the effective four-dimensional action
\begin{eqnarray}
\label{bdaction}
S_4 = \int d^4x \sqrt{-f} e^{\beta} \left[ R^{(f)} -
\frac{2}{q^2} \left( \nabla \beta \right)^2 - V \right]
\\
V(\beta ) \equiv -2c_1 e^{-2\beta} + \frac{\sigma_0^2}{2} 
e^{-[2+(4/q^2)]  \beta}  ,
\end{eqnarray}
where $R^{(f)}$ is the Ricci curvature scalar of the metric 
$f_{\mu\nu}$ and $f\equiv {\rm det} f_{\mu\nu}$. 
The action (\ref{bdaction}) represents an effective Brans--Dicke 
scalar--tensor theory of gravity with a constant dilaton-graviton coupling 
parameter given by $\omega = 2/q^2$. It should be emphasized that 
no a priori assumption has been made regarding 
the form of the world--volume metric, $f_{\mu\nu}$. 
Thus, we have found a general class of warped, five--dimensional 
geometries of the form (\ref{5dmetric}), where the world--volume
metric satisfies the field equations derived from the effective 
action (\ref{bdaction}). Models of this type, both with and 
without an axion field, but with vanishing 
separation constants $c_i=0$,  have been analyzed previously 
in a number of settings \cite{lukas,specific,lidsey}. 

\subsection{Warp Factor}

It only remains to solve Eqs. (\ref{c1define})--(\ref{c3define})
for the warp factor $H(y)$. However, these equations 
are independent of the world--volume coordinates, so the warp factor 
takes an identical form to that for the 
spatially flat FRW world--volume. Its precise form is 
determined by the values of the parameters $\{ q, V_0, c_1 \}$ and, 
since our interest here is in the nature of the world--volume metric, we refer 
the reader to Refs. \cite{koyama,mc} for further details. 
As an example, however, consider the case of a two--brane 
scenario supported by a negative bulk potential, 
$V_0<0$, where $q^2< 2/3$ and $c_1<0$. It follows from 
Eqs. (\ref{c1define})--(\ref{c3define}) that 
\begin{equation}
\label{warpH}
H= \sqrt{\frac{|V_0|}{2c_2}} \sinh \left( D|y-y_0| \right) , 
\qquad 
D \equiv \sqrt{\frac{c_2(3q^2-2)^2}{3(8-3q^2)}}  ,
\end{equation}
where the integration constant $y_0$ is 
specified such that the branes are located at $y= (0,\pi)$, i.e., 
$H(0) = H(\pi ) = 1$. This implies that the brane tensions are given by  
\begin{equation}
\mu_1 = - \sqrt{ \frac{24(2c_2+|V_0|)}{8-3q^2}} , \qquad 
\mu_1=-\mu_2  ,
\end{equation}
where the second relation follows as a consequence of the 
${\rm Z}_2$ reflection symmetry.
 
\subsection{Cosmic No Hair on the World--Volume}

We are now able to deduce a cosmic no hair 
theorem for this class of braneworlds when the world--volume 
represents a spatially 
homogeneous but anisotropic Bianchi spacetime. (A Bianchi metric admits 
three--dimensional, space--like hypersurfaces on which a three--parameter 
Lie group of isometries acts simply transitively). 
To proceed, we note that action (\ref{bdaction}) 
may be transformed into the Einstein--Hilbert action for a  
minimally coupled, self--interacting scalar field 
by the conformal transformation
\begin{equation}
\label{conformaltrans}
\tilde{f}_{\mu\nu} = \Omega^2 f_{\mu\nu} , \qquad \Omega^2 
\equiv e^{\beta}
\end{equation}
and field redefinition
\begin{equation}
\label{chidef}
\chi \equiv \sqrt{3 +\frac{4}{q^2}} \beta  . 
\end{equation}
It follows that 
\begin{equation}
\label{einsteinaction}
\tilde{S}=\int d^4 x \sqrt{-\tilde{f}} \left[ \tilde{R}
-\frac{1}{2} \left( \tilde{\nabla} \chi \right)^2
- \tilde{V} (\chi ) \right]  ,
\end{equation}
where
\begin{equation}
\label{effpot}
\tilde{V} = -2c_1e^{-\lambda \chi} + \frac{\sigma_0^2}{2}
e^{-3 \chi /\lambda} , \qquad \lambda \equiv 
\frac{3|q|}{\sqrt{4+3q^2}}   .
\end{equation}

The effective potential (\ref{effpot}) 
contains two contributions, one from the axion field and 
the other from the non--trivial separation constant $c_1$. When $c_1<0$, 
the potential is positive--definite and contains no
turning points. This implies that $\chi \rightarrow + \infty$ 
at late--times for any ever--expanding cosmology.
(For the case where $c_1 >0$, the potential exhibits a 
minimum but its value is negative at this point).  
Since the contribution to the potential sourced by the axion field 
is steeper, the self--interactions of the 
field become dominated at late--times by the contribution arising from the 
separation constant. Now, the conformal transformation 
(\ref{conformaltrans})--(\ref{chidef}) is well--defined
for the class of spatially homogeneous and anisotropic Bianchi metrics and,  
moreover, the two metrics in Eq. (\ref{conformaltrans}) correspond 
to the same Bianchi type if the radion field $\beta$ is constant on the 
surfaces of homogeneity, i.e., the Bianchi type is invariant under the 
conformal transformation. This implies that 
the known results from four--dimensional 
general relativity coupled to an exponential potential 
can be carried over directly to this
braneworld scenario \cite{km}. 
We are therefore led to the following cosmic 
no hair theorem for braneworlds with a separable metric of the form 
(\ref{5dmetric}) supported by a bulk exponential potential:
{\em for $q^2<2/3$, 
all initially expanding, spatially homogeneous Bianchi world--volumes 
(except for a subclass of Bianchi type IX models that recollapse) 
isotropize in the future toward the power--law  
inflationary, spatially flat FRW metric 
$ds^2 =-d\tau^2 +\tau^{4/(3q^2)}\delta_{ij}dx^idx^j$}. (The 
form of the world--volume metric follows after conformally 
transforming back to the original frame (\ref{bdaction}).)

Thus, the scaling solution (\ref{scalingsol}) represents a
late--time attractor for spatially homogeneous models when the separable ansatz 
(\ref{5dmetric}) applies and $q^2<2/3$. 
In the following Section, we investigate the implications 
of relaxing this assumption on the form of the bulk metric. 
 
\section{Inhomogeneities in the Bulk and the Hamilton--Jacobi Formalism}

In the previous Section we deduced a cosmic no hair result for 
braneworlds satisfying the separable ansatz (\ref{5dmetric})
with a spatially homogeneous world--volume. 
We now wish to provide evidence that the scaling metric 
(\ref{scalingsol}) also represents an attractor under more 
general inhomogeneous settings. This involves a study of 
the five--dimensional bulk solutions when the separable ansatz is relaxed.  
A powerful framework for solving the Einstein 
field equations sourced by a self--interacting scalar field is 
provided by the Hamilton--Jacobi (HJ) formalism of 
general relativity. We 
briefly review this formalism in a five--dimensional context in the 
following Subsection and then proceed to investigate 
the evolution of the bulk metric. 

\subsection{Hamilton-Jacobi Equation}

We consider the five--dimensional sector of action (\ref{action5})
with vanishing axion field and investigate bulk metrics 
in the Arnowitt-Deser-Misner (ADM) form \cite{mtw}
\begin{equation}
\label{defgamma}
ds^2 = (N^2 + \gamma_{\mu\nu} N^{\mu}N^{\nu} ) 
dy^2 +2N_{\mu} dydx^{\mu} + 
\gamma_{\mu\nu} dx^{\mu} dx^{\nu}   ,
\end{equation}
where $\gamma_{\mu\nu}= \gamma_{\mu\nu} (x^{\rho}, y)$.
Rewriting action (\ref{action5}) in a Hamiltonian form and 
varying with respect to the lapse and shift functions, $N$ and $N^{\mu}$, 
yields the Hamiltonian and momentum constraints, whereas 
the equations of motion for $\gamma_{\mu\nu}$ and $\varphi$ 
follow by varying the action with respect to their conjugate momenta, 
$\pi^{\mu\nu}$ and $\pi^{\varphi}$: 
\begin{eqnarray}
\label{gammaeom}
\partial_y \gamma_{\mu\nu} - \nabla_{\nu} N_{\mu} - \nabla_{\mu} N_{\nu}
= -2\frac{N}{\sqrt{-\gamma}} \pi^{\lambda\rho} 
\left( \gamma_{\mu\rho}\gamma_{\nu\lambda}- 
\frac{1}{3} \gamma_{\mu\nu}\gamma_{\lambda\rho} \right)
\\
\label{eomPhi}
\partial_y \varphi - N^{\mu} \nabla_{\mu} \varphi = - \frac{N}{\sqrt{-\gamma}} 
\pi^{\varphi}   .
\end{eqnarray}
The evolution equations for the momenta are automatically 
solved \cite{mtw} by defining 
\begin{equation}
\label{defmomenta}
\pi^{\mu\nu} = \frac{\delta S}{\delta \gamma_{\mu\nu}} , \qquad 
\pi^{\varphi}= \frac{\delta S}{\delta \varphi}
\end{equation}
and requiring that 
these satisfy the Hamiltonian and momentum constraints when 
the equations of motion (\ref{gammaeom})--(\ref{eomPhi}) are satisfied. 
The functional $S$ represents the generating functional     
of the HJ equation. The momentum constraint implies that 
this functional should be diffeomorphism invariant \cite{momsolve} and  
the HJ equation then represents the Hamiltonian constraint. 
This is a hyperbolic, functional partial 
differential equation for $S$ and may be expressed in the form  
\begin{equation}
\label{HJequation}
\{ S,S \} = {\cal{L}}_4   ,
\end{equation}
where we have defined the bracket: 
\begin{equation}
\label{defbracket}
\{ S,S \} \equiv - \frac{1}{\sqrt{-\gamma}} 
\frac{\delta S}{\delta \gamma_{\mu\nu}} \frac{\delta S}{\delta 
\gamma_{\lambda\rho}} \left( \gamma_{\mu\rho}\gamma_{\nu\lambda} - 
\frac{1}{3} \gamma_{\mu\nu}\gamma_{\lambda\rho} \right)
- \frac{1}{2\sqrt{-\gamma}} \left( \frac{\delta S}{\delta \varphi} \right)^2
\end{equation}
and  
\begin{equation}
\label{defL4}
{\cal{L}}_4 \equiv \sqrt{-\gamma} R (\gamma )- \frac{1}{2} \sqrt{-\gamma}
\gamma^{\mu\nu} \nabla_{\mu}\varphi \nabla_{\nu} \varphi - \sqrt{-\gamma}
V(\varphi)   .
\end{equation}

Eqs. (\ref{gammaeom})--(\ref{defmomenta}) yield the full set of 
evolution equations given a solution to the HJ equation (\ref{HJequation}).  
The key idea of the spatial gradient expansion method \cite{sal1,sal2,sal3}
is to derive an order--by--order solution of the HJ equation by 
expanding the generating functional in a series $S=\sum_{n=0}^{\infty}
S^{(2n)}$, where $2n$ represents the number of spatial gradients in 
$S^{(2n)}$. The HJ equation is also expanded in spatial 
gradients, ${\cal{H}}=\sum_{n=0}^{\infty} {\cal{H}}^{(2n)} =0$, and 
is then required to vanish at each order in $n$. 
For $n=(0,2,4)$, this leads to the constraints: 
\begin{eqnarray}
\label{HJ0}
\{ S^{(0)} , S^{(0)} \} = - \sqrt{-\gamma} V(\varphi)
\\
\label{HJ2}
2\{ S^{(0)} , S^{(2)} \} = \sqrt{-\gamma} R - \frac{1}{2} 
\sqrt{-\gamma} \left( \nabla \varphi \right)^2 
\\
\label{HJ4}
2 \{ S^{(0)}, S^{(4)} \} + \{ S^{(2)} , S^{(2)} \} =0  .
\end{eqnarray}

The zero--order equation (\ref{HJ0}) is solved by \cite{sal1}
\begin{equation}
\label{action0ansatz}
S^{(0)} = - \int d^4x \sqrt{-\gamma} W(\varphi)   ,
\end{equation}
where the function $W(\varphi )$ is a solution to 
the ordinary differential equation (ODE): 
\begin{equation}
\label{HJzero}
\frac{1}{2} \left( \frac{d W}{d \varphi} \right)^2  
-\frac{1}{3} W^2 = V  .
\end{equation}
Each solution to Eq. (\ref{HJzero}) is characterized 
by the value of a single parameter field, $\tilde{\varphi}$, such 
that $W=W(\varphi , \tilde{\varphi})$. Differentiating 
(\ref{HJzero}) with respect to $\tilde{\varphi}$ then implies that 
the solution to Eq. (\ref{HJzero}) can be expressed as \cite{sal2} 
\begin{equation}
\label{Wquad}
W= \frac{3}{2} \frac{\partial W}{\partial \varphi} 
\frac{\partial}{\partial \varphi} \ln \left( \frac{\partial W}{\partial
\tilde{\varphi}} \right)  .
\end{equation}

The second--order equation (\ref{HJ2}) is solved by substituting 
the ansatz 
\begin{equation}
\label{action2ansatz}
S^{(2)} = \int d^4x \sqrt{-\gamma} \left[ J(\varphi)R - \frac{1}{2}K(\varphi)
\left( \nabla \varphi \right)^2 \right]
\end{equation}
for the functions $J(\varphi )$ and $K(\varphi )$ 
and requiring the coefficients of the terms involving 
$R$, $\square \varphi$ and $(\nabla \varphi )^2$ to vanish identically. 
This results in three coupled ODEs \cite{sal2}:  
\begin{eqnarray}
\label{seconda}
\frac{dW}{d\varphi} \frac{dJ}{d\varphi} - \frac{1}{3} WJ =1
\\
\label{secondb}
W\frac{dJ}{d\varphi} + K\frac{dW}{d\varphi} =0
\\
\label{secondc}
2W\frac{d^2J}{d\varphi^2} + \frac{dK}{d\varphi}\frac{dW}{d\varphi} 
+ \frac{1}{3} WK =-1  .
\end{eqnarray}
Eqs. (\ref{seconda})--(\ref{secondc}) are not independent and 
Eq. (\ref{secondc}) follows after 
differentiating Eqs. (\ref{seconda}) and (\ref{secondb}) with 
respect to $\varphi$. Moreover, Eq. (\ref{seconda}) may be solved in terms of 
an integrating factor once a solution to Eq. (\ref{HJzero}) has been 
found. It follows, after substitution of the solution (\ref{Wquad})
into Eq. (\ref{seconda}), that  
\begin{equation}
\label{Jquad}
J = \left( \frac{\partial W}{\partial \tilde{\varphi}} \right)^{1/2}
\int d \varphi \, \left( \frac{\partial W}{\partial \varphi} \right)^{-1} 
\left( \frac{\partial W}{\partial \tilde{\varphi}} \right)^{-1/2}  .
\end{equation}

In principle, therefore, the HJ equation can be solved perturbatively 
once a solution to the zero--order equation (\ref{HJzero}) 
has been found. We consider the case of an exponential potential 
in the following Subsections. 

\subsection{Solution for a Bulk Exponential Potential}

In general, Eq. (\ref{HJzero}) may be viewed as a first--order ODE 
with a `time' variable $\varphi$ \cite{lidseywaga}. 
By defining a new dependent 
variable \footnote{We assume implicitly and 
without loss of generality that $dW/d\varphi <0$.}  
\begin{equation}
\label{defY}
Y \equiv - \sqrt{\frac{2}{3}}\frac{W}{dW/d\varphi} , \qquad \frac{1}{Y^2} = 
1 + 3\frac{V}{W^2}  ,
\end{equation}
this ODE may be expressed in the form of a first--order 
Abel equation: 
\begin{equation}
\label{abel}
\sqrt{\frac{3}{2}}\frac{dY}{d\varphi} =  
\sqrt{\frac{3}{8}} \frac{d \ln V}{d\varphi} 
Y^3 + Y^2 - \sqrt{\frac{3}{8}} \frac{d \ln V}{d\varphi} 
Y -1  .
\end{equation}

For a scalar field with a negative exponential potential, 
$V=V_0 \exp (-q \varphi )$ with $V_0<0$ and $q>0$, 
it proves convenient to define a further variable 
$Y \equiv \Theta^{-1/2}$. Since $\Theta$ is bounded such that 
$0 \le \Theta^2 \le 1$, this implies that  
Eq. (\ref{abel}) can then be written as a one--dimensional
non--linear dynamical system: 
\begin{equation}
\frac{d\Theta}{d\varphi} = \sqrt{\frac{8}{3}} (\Theta -1) 
\left( \sqrt{\Theta} -s \right)  ,
\end{equation}
where $s \equiv \sqrt{3/8}q$. The equilibrium points for this system are at 
$\Theta_{\rm eqm} = s^2$ and $\Theta_{\rm eqm} =1$, respectively. 
A stability analysis in the neighbourhood of these points implies 
that the former is stable for $q < \sqrt{8/3}$ and unstable for 
$q>\sqrt{8/3}$, whereas the latter equilibrium point is unstable 
for $q<\sqrt{8/3}$ and stable for $q>\sqrt{8/3}$. 
Since $\Theta$ is bounded, these points represent 
the global attractor and repellor in the phase space. The point 
$\Theta_{\rm eqm}=1$ corresponds to the limit where the 
potential energy of the field is dynamically negligible, whereas 
$\Theta_{\rm eqm} = s^2$ represents scaling behaviour. We therefore 
focus in the remainder of this Section on the region of parameter space 
where $q < \sqrt{8/3}$.
Moreover, the general solution to Eq. (\ref{abel}) is given by  
\begin{equation}
\frac{(1+Y)^{s-1}(1-Y)^{s+1}}{(1-sY)^{2s}} = \exp \left[ 
\sqrt{\frac{8}{3}}(1-s^2) (\varphi -\varphi_m ) \right]  ,
\end{equation}
where $\varphi_m$ is an integration constant, and it follows that 
the stable equilibrium point corresponds to 
$\varphi \rightarrow +\infty$. (This will be important 
when calculating the asymptotic form of the third--order 
metric $\gamma^{(3)}_{\mu\nu}$ in Section IIID). 

The attractor solution to the zero--order HJ equation now
follows immediately from Eq. (\ref{defY}): 
\begin{equation}
\label{Wattract}
W(\varphi)   = W_0 e^{-q\varphi /2} , \qquad W_0 = 
\pm  \sqrt{\frac{24V_0}{3q^2-8}}  .
\end{equation}
Eqs. (\ref{seconda}) and (\ref{secondb}) are then solved by
\begin{equation}
\label{Jattract}
K(\varphi) = J(\varphi) , \quad J(\varphi ) = J_0 e^{q\varphi /2} , \quad 
W_0J_0 = -\frac{12}{3q^2+4}
\end{equation}
and the generating functional to second--order in spatial gradients 
is therefore given by 
\begin{equation}
\label{action2attract}
S^{(0)} +S^{(2)}  = \int d^4 x \sqrt{-\gamma} \left[ 
J_0e^{q\varphi /2} \left( R - \frac{1}{2} \left( \nabla \varphi \right)^2
\right) - W_0  e^{-q \varphi /2} \right]  .
\end{equation}
Modulo trivial rescalings and a field redefinition $\varphi 
\rightarrow 2\beta /q$, 
the coupling between the scalar and tensor degrees of 
freedom in Eq. (\ref{action2attract}) is {\em precisely} 
the same as the coupling in 
the effective four--dimensional action (\ref{bdaction}), 
i.e., it is of the Brans--Dicke form where $\omega=2/q^2$. The qualitative 
forms of the potential terms are also identical. 

\subsection{Fourth-Order Hamiltonian}

We now solve the fourth--order equation (\ref{HJ4}). 
Substitution of the variations of $S^{(0)}$ yields 
the first--order functional differential equation 
\begin{equation}
\label{RGflow}
W \frac{\delta S^{(4)}}{\delta \gamma_{\mu\nu}} \gamma_{\mu\nu}
- 3 \frac{dW}{d\varphi} \frac{\delta S^{(4)}}{\delta \varphi} = 3 \{ S^{(2)} , 
S^{(2)} \}
\end{equation}
and this equation can be solved by employing the 
conformal transformation technique of Ref. \cite{sal3}. 
This involves defining the set of new variables: 
\begin{equation}
\label{defu}
\gamma_{\mu\nu} \equiv \Omega^2 (u) k_{\mu\nu} , \qquad 
\frac{\partial \Omega}{\partial u} = \frac{W}{2} \Omega , \qquad 
u \equiv -\frac{1}{3} \int d\varphi \left( \frac{dW}{d\varphi} \right)^{-1}
\end{equation}
and transforming Eq. (\ref{RGflow}) into the form 
\begin{equation}
\label{functionalde}
\frac{\delta S^{(4)}}{\delta u} = 3 \{ S^{(2)}, S^{(2)} \}  .
\end{equation}
Eq. (\ref{functionalde}) then admits a solution in terms of the 
line integral 
\begin{equation}
\label{formalintegral}
S^{(4)} = 3 \int_0^u du' \, \int d^4x \, {\cal{R}}^{(4)} [ 
u'(x) , k_{\mu\nu} (x) ]  ,
\end{equation}
where ${\cal{R}}^{(4)} \equiv \{ S^{(2)} , S^{(2)} \}$ is to be viewed 
as a functional of $u'(x)$ and the conformal metric $k_{\mu\nu}$.
The integral (\ref{formalintegral}) may be evaluated  by choosing 
a straight line path such that \cite{sal3}
\begin{equation}
\label{lineintegral}
u'(x) = r u (x), \qquad du' (x) = u(x) dr, 
\end{equation}
where $r$ is a real parameter taking values in the range $0\le r \le 1$. 
Each term in ${\cal{R}}^{(4)}$ depends quadratically 
on $r$ and, consequently, the integral over $u$ reduces to  
performing the trivial integration 
$\int^1_0 r^2dr$. The solution to Eq. (\ref{HJ4}) is therefore 
given by 
\begin{equation}
\label{integralsolution}
S^{(4)} = \int d^4 x\, u \{ S^{(2)}, S^{(2)} \}  .
\end{equation}

The explicit form of the fourth--order contribution 
to the generating functional then follows after substitution of 
the second--order contribution $S^{(2)}$. In the case of the 
attractor solution (\ref{Wattract})--(\ref{Jattract}) 
for the exponential potential, it follows from Eq. (\ref{defu})
that $u=4/(3q^2W)$. Hence, we need only substitute the second--order 
term, Eq. (\ref{action2attract}),  
into the integral (\ref{integralsolution}) and integrate by parts 
where appropriate. We find, after some algebra, that 
\begin{eqnarray}
\label{nonlocalaction}
S^{(4)}  =-\frac{4J_0^2}{3q^2 W_0} \int d^4 x \sqrt{-\gamma} 
e^{3q\varphi /2} \left[ R_{\mu\nu}R^{\mu\nu} - 
\left( \frac{1}{3} - \frac{q^2}{8} \right) R^2 
+ \frac{1}{2} \left( \square \varphi \right)^2 + \frac{q}{2} R \square \varphi + 
\left( \frac{1}{3} - \frac{3}{8} q^2 \right) R \left( \nabla \varphi \right)^2 
\right. 
\nonumber \\
\left. - \left( 1-\frac{3q^2}{4} \right) R_{\mu\nu} \nabla^{\mu} \varphi 
\nabla^{\nu} \varphi - \frac{q}{4} \left( 1-\frac{3q^2}{4} \right) \square \varphi 
\left( \nabla \varphi \right)^2 \right.  
\nonumber \\
\left. 
+ \left( \frac{1}{6} - \frac{7q^2}{32} + \frac{3q^4}{32} \right) \left( 
\nabla \varphi \right)^4 \right]  .
\end{eqnarray}

\subsection{Evolution of the world--volume metric} 

We are now able to evaluate the evolution of the metric $\gamma_{\mu\nu}$ 
up to third--order in derivatives. To proceed, 
we specify the gauge such that 
the shift function vanishes, $N^{\mu}=0$, and further 
assume that the scalar field is constant on surfaces of constant $y$, 
i.e., $\varphi =\varphi (y)$. Moreover, we identify  
the value of the scalar field as the `time' parameter representing 
evolution in the fifth dimension. This is equivalent to choosing the lapse 
function to be   
\begin{equation}
\label{lapse}
\frac{1}{N} = \frac{\partial W}{\partial \varphi} - 
\frac{\partial J}{\partial \varphi} R ,
\end{equation}
as follows directly from Eq. (\ref{eomPhi}).

The evolution of the metric
$\gamma_{\mu\nu}$ to first--order is determined by truncating the generating 
functional at the lowest--order term, $S=S^{(0)}$, 
and setting the lapse $N^{-1} = \partial W/\partial \varphi$ in 
Eq. (\ref{gammaeom}) \cite{sal2}. 
Integrating (\ref{gammaeom}) then yields 
\begin{equation}
\label{gamma1}
\gamma_{\mu\nu}^{(1)} = \left( \frac{\partial W}{\partial \tilde{\varphi}}
\right)^{-1/2} h_{\mu\nu} (x^{\rho})  ,
\end{equation}
where we have employed expression (\ref{Wquad}) and 
the conformal metric, $h_{\mu\nu}(x)$, 
is independent of the fifth coordinate. 

The evolution of the metric to third--order, 
on the other hand, is determined by truncating the 
expansion of the generating functional at $n=1$. After substituting 
the variation of $S^{(2)}$, as determined from 
Eq. (\ref{action2ansatz}),
into the evolution equation (\ref{gammaeom}), it follows that 
\begin{equation}
\label{evolutionsofar}
\frac{\partial_y \gamma_{\mu\nu}}{N} = -\frac{1}{3} W \gamma_{\mu\nu}
+ 2J \left( R_{\mu\nu} - \frac{1}{6} R\gamma_{\mu\nu} \right)  ,
\end{equation} 
where the second term on the right hand side is evaluated 
with the first--order (long--wavelength) metric (\ref{gamma1}) and 
the other two terms contain contributions from the first-- and third--order 
metrics. The third--order metric is determined by integrating 
(\ref{evolutionsofar}) after substitution of Eq. (\ref{lapse}), 
where the substitution is done in such a way that only terms up to 
second--order in spatial gradients are retained 
and first--order results are substituted into second--order terms
\cite{sal2}. It is found that  
\begin{eqnarray}
\label{gamma3}
\gamma^{(3)}_{\mu\nu} (x^{\rho}, \varphi )
= \left( \frac{\partial W}{\partial \tilde{\varphi}}
\right)^{-1/2} h_{\mu\nu} 
- \frac{1}{3} \left( \frac{\partial W}{\partial \tilde{\varphi}}
\right)^{-1/2} \int d \varphi' \, W \frac{\partial J}{\partial \varphi'}
\left( \frac{\partial W}{\partial \varphi'} \right)^{-2} 
\left( \frac{\partial W}{\partial \tilde{\varphi}} \right)^{1/2}
R^{(h)} h_{\mu\nu} 
\nonumber \\
+ 2 \left( \frac{\partial W}{\partial \tilde{\varphi}}
\right)^{-1/2} \int d \varphi' \, J \left( \frac{\partial W}{\partial \varphi'}
\right)^{-1}  \left( \frac{\partial W}{\partial \tilde{\varphi}} 
\right)^{1/2} \left[ R^{(h)}_{\mu\nu} - \frac{1}{6} 
R^{(h)} h_{\mu\nu} \right]  ,
\end{eqnarray}
where $R^{(h)}_{\mu\nu}$ and $R^{(h)}$ are the Ricci tensor and 
curvature scalar, respectively, of the conformal metric $h_{\mu\nu}$.

It follows from Eq. (\ref{gamma3}) 
that the evolution of the metric at this order is determined, 
at least in principle, once the zero--order HJ equation (\ref{HJzero}) has 
been solved. To determine the asymptotic behaviour 
of the world--volume metric, therefore, we may substitute the 
attractor solution (\ref{Wattract}) into Eqs. (\ref{Wquad}) and 
(\ref{Jquad}). This is equivalent to substituting 
$\partial W /\partial \tilde{\varphi} =e^{-(4/3q)\varphi}$ into Eqs. 
(\ref{gamma1}) and (\ref{gamma3}) and we deduce that 
\begin{eqnarray}
\label{gamma1asymp}
\gamma^{(1)}_{\mu\nu} = e^{(2/3q)\varphi} k_{\mu\nu}
\\
\label{gamma3asymp}
\gamma_{\mu\nu}^{(3)} = e^{(2/3q) \varphi} k_{\mu\nu}
- \frac{12J_0}{W_0(3q^2-2)} e^{q\varphi} R^{(k)}_{\mu\nu}  ,
\end{eqnarray}
where $k_{\mu\nu}$ is directly proportional 
to $h_{\mu\nu}$. 
Hence,  since $\varphi \rightarrow \infty$ corresponds to 
the attractor in this 
model when $q^2 < 8/3$, we conclude that the first--order 
term increases {\em more rapidly} than the third--order term
for $q^2 < 2/3$. As a result, 
the metric approaches the first--order separable metric 
(\ref{gamma1asymp}) for this region of parameter space. 
This is precisely the upper limit 
on the value of the coupling parameter for which the cosmic 
no hair theorem of Section II applies. Moreover, comparison with 
the exact braneworld (\ref{5dmetric}) and (\ref{offdiagsolve}) 
implies that the world--volume sector of the bulk
solution (\ref{5dmetric}) can be expressed in exactly  
the same form, $\gamma_{\mu\nu} = e^{(2/3q) \varphi} f_{\mu\nu} (x)$, 
as that of the first--order metric (\ref{gamma1asymp}). 

\section{Discussion and A Note on Holography}

In this paper, we have focused on various aspects of 
braneworld cosmology where the bulk gravitational action 
contains a scalar dilaton field with an exponential 
self--interaction potential with coupling parameter $q$. 
A massless axion field coupled to the dilaton was also included in 
the bulk action. We have found a general class of braneworlds, where 
the world--volume metric represents any solution 
to a four--dimensional scalar--tensor theory of gravity where the coupling 
between the spin--0 and spin--2 fields 
takes the value $\omega = 2/q^2$. The axion field 
generates a potential for the dilaton in four dimensions. 
This generalizes the results
of \cite{koyama,mc,kt} to an arbitrary world--volume metric in the presence 
of a bulk axion field and extends the results of 
\cite{lidsey,specific} to the case 
where the four--dimensional dilaton has a non--trivial potential. 

We have argued, from both the world--volume and bulk perspectives, 
that the spatially flat FRW scaling solution 
(\ref{scalingsol}) represents an asymptotic attractor 
in a wide variety of settings when the constraint $q^2<2/3$ 
is satisfied. Specifically, we have derived a `cosmic no hair' 
theorem for the class of spatially homogeneous 
Bianchi world--volumes. We also applied the Hamilton--Jacobi framework 
to five--dimensional general relativity up to third--order in 
metric derivatives and found that when $q^2<2/3$, the 
first--order (separable) bulk metric dominates the third--order contributions 
as the attractor is approached. Moreover, it is striking that the zero-- and
second--order contributions to the generating functional 
of the HJ equation take the same form as the effective 
Brans--Dicke action that determines the world--volume metric 
$f_{\mu\nu}$ in the separable bulk solution. We have also presented 
the first derivation of the fourth--order 
contribution to the generating functional for five--dimensional 
Einstein gravity coupled to an exponential scalar field potential. 

An alternative approach to solving the bulk Einstein equations is to employ a 
gradient (low-energy) expansion technique directly at the level of 
the field equations \cite{effective,effective1}. The two-brane 
scenario with a bulk exponential potential was recently studied in this context 
by Leeper {\em et al.} \cite{leeper}, who found an approximate solution 
to the field equations where a perturbation 
in the position of one of the branes induces a perturbation 
in the bulk metric away from the separable ansatz (\ref{scalingsol}). 
It was found that the system is stable to such a perturbation and 
that the bulk rapidly tends to its unperturbed form. Such an analysis 
differs from that of the present work and 
was restricted to first-order perturbations and it would clearly 
be of interest to extend the analysis to higher-order. 

Although our primary interest has focused on bulk gravitational 
issues, our work also overlaps with recent developments 
in holographic approaches to cosmology and 
we now conclude with a discussion on these issues. 
The AdS/CFT correspondence \cite{adscft} states that gravity on 
$(d+1)$--dimensional anti-de Sitter (AdS) space admits a dual description 
in terms of a conformal field theory (CFT) on the 
$d$--dimensional boundary. (For a review, see, \cite{adsreview}).
Within this context, the radial bulk coordinate is identified 
as a renormalization group (RG) flow parameter (energy scale) 
of the dual field theory, 
such that the evolution of the bulk fields along the radial direction induces 
non--vanishing $\beta$--functions in the dual theory
\cite{flowrefs,rgflow,boer}. As is well known, however,
the supergravity action diverges when the AdS boundary is taken to infinity.
de Boer, Verlinde and Verlinde \cite{boer} have advocated a 
method of handling such divergences within a holographic RG approach 
based on the HJ formalism. The effective action for the gauge theory, 
$\Gamma$, is related to the bulk classical action, $S$, 
by $S=S_{\rm loc} + \Gamma$, where $S$ is evaluated on a solution to 
the bulk field equations (with appropriate boundary conditions),  
$S_{\rm loc}$ represents the divergent terms (that are no higher than
second--order in derivatives) and $\Gamma$ contains the
higher--order, non--local contributions. 
A primary motivation for such an approach is that equation 
(\ref{HJ4}), determining the fourth--order contribution to the 
generating functional of the HJ equation, takes the form of 
a `Callan-Symanzik' equation \cite{boer}: 
\begin{equation}
\label{CS}
\gamma^{\mu\nu} \frac{\delta \Gamma}{\delta \gamma^{\mu\nu}} = 
\beta (\varphi ) \frac{\delta \Gamma}{\delta \varphi} + \frac{3}{W} 
\{ S_{\rm loc} , S_{\rm loc} \}  ,
\end{equation}
when we identify $S^{(4)}=\Gamma$, $S^{(2)} = S_{\rm loc}$, and 
$\beta \equiv 3d \ln W /d \varphi$ with the $\beta$--function of 
the dual theory. The left--hand side 
of Eq. (\ref{CS}) is precisely the conformal anomaly of the gauge theory 
\cite{verify,sergei,noz,kiritsis}. 

Within the context of the braneworld paradigm, this provides strong 
motivation for interpreting the braneworld as a cut--off, strongly 
coupled conformal gauge theory coupled to four--dimensional 
gravity with a dual action 
$\tilde{S}  = S_{\rm loc} + \Gamma + S_{\rm brane}$, 
where $S_{\rm brane}$ represents the brane 
\cite{brave,noz,kiritsis}. Thus, the dual effective action 
in the presence of a bulk dilaton scalar field would be determined 
in this context from Eqs. (\ref{HJ0})--(\ref{secondb}). 

The bulk solutions found in Section II represent domain wall backgrounds, 
but these are not asymptotically AdS since the exponential potential 
does not contain a global minimum. Nonetheless, 
the above discussion should apply to 
any model where the bulk potential has an approximately 
exponential form over some finite range of scalar field values and, 
in this case, the fourth--order contribution to 
the HJ generating functional that we have derived in Eq. (\ref{nonlocalaction}) 
may then be identified as the effective action for the conformal anomaly
in this regime. It would clearly be of interest to establish the necessary 
conditions for inflation to arise from such a 
contribution, since its ultra--violet nature
implies that it may play a dominant role in the 
very early universe. On the other hand, we have found that 
the low--energy limit of the dual effective action 
can isotropize the braneworld if the logarithmic derivative of 
the (negative) bulk potential is sufficiently flat, i.e., the 
$\beta$--function of the gauge theory is sufficiently small.  

Finally, the bulk solutions we have investigated in 
the present work may also be relevant to 
the proposed domain--wall/quantum field theory (DW/QFT) correspondence
\cite{dwqft}, which exploits the fact that AdS space (in horospherical 
coordinates) represents a special case of a domain--wall background. 
This suggests -- in view of the dualities that relate all 
brane backgrounds -- that the AdS/CFT correspondence can be 
extended to that of an ordinary QFT living on the boundary of the domain wall. 
More specifically, the metric of AdS space in horospherical coordinates
with radius of curvature $\ell$ is given by 
$ds^2 = \ell^2 (du^2/u^2) + (u^2/\ell^2 ) \eta_{\mu\nu}dx^{\mu}
dx^{\nu}$, where $\eta_{\mu\nu}$ is the metric for flat space. 
In general, the bulk metric we have considered 
as an ansatz in Eq. (\ref{5dmetric}) can not be expressed in this form. 
However, when the separable condition (\ref{offdiagsolve})
is satisfied, a conformal transformation 
\begin{equation}
\label{dualconftran}
ds^2_{\rm dual} = e^{-q\varphi} d\hat{s}^2_5
\end{equation}
on the metric (\ref{5dmetric}) results in the `dual' metric  
\begin{equation}
\label{dualmetric}
ds^2_{\rm dual} = H^{-2}(y) \left( e^{-2\beta} f_{\mu\nu} dx^{\mu}
dx^{\nu} + dy^2 \right) .
\end{equation}
Comparison with the warp factor (\ref{warpH}), for example, then implies  
that in the limit $y \rightarrow y_0$, the metric 
(\ref{dualmetric}) can indeed be expressed in the horospherical
AdS form (after a trivial rescaling of the coordinates) by identifying 
$(y-y_0) \propto u^{-1}$. 

In the DW/QFT correspondence, the horospherical coordinate 
$u$ is identified as the energy scale of the dual theory, with $u=\infty$ 
corresponding to the AdS boundary \cite{dwqft}. 
The dual frame (\ref{dualmetric}) therefore
provides the natural context for discussing the DW/QFT correspondence. 
Moreover, performing the conformal transformation 
(\ref{dualconftran}) on the  bulk action (\ref{action5})--(\ref{potential5}) 
results in a scalar--tensor gravity theory:
\begin{equation}
\label{dualstaction}
S_{\rm dual} = \int d^5 x \sqrt{-g} e^{-\varphi_{\rm dual}}
\left[ R - \omega_{\rm dual} \left( \nabla \varphi_{\rm dual}
\right)^2 - V_0 \right]  ,
\end{equation}
where
\begin{equation}
\label{redefine}
\varphi_{\rm dual} \equiv -\frac{3q}{2}  \varphi , \qquad 
\omega_{\rm dual} \equiv \frac{2}{9q^2} (1-6q^2)  .
\end{equation}
It is of interest to note that the critical value we have 
identified for the attractor scaling solution, $q^2=2/3$, 
corresponds precisely to the coupling 
that arises in the dilaton--graviton 
sector of the string effective action, $\omega_{\rm dual} =-1$.
In conclusion, therefore, we anticipate that the class of braneworlds 
we have found will play a key role in developing the 
holographic approach to braneworld cosmology in terms of the 
AdS/CFT and DW/QFT correspondences. 

\section*{Acknowledgments}
DS is supported by PPARC. We thank A. Coley, K. Koyama, E. Leeper and 
S. Odintsov for helpful comments and discussions.

\end{document}